\documentclass[a4paper,11pt]{article}
\pdfoutput=1

\usepackage[a4paper,
  left=2.5cm, right=2.5cm,
  top= 3cm, bottom=4cm]{geometry}

\usepackage{amsmath}
\numberwithin{equation}{section}
\usepackage{mathtools}
\usepackage{oldgerm}
\usepackage{amssymb}
\usepackage{bbm}
\usepackage{framed}
\usepackage{xcolor}

\usepackage{graphicx}
\usepackage{subcaption}
\usepackage{tikz-cd} 
\usetikzlibrary{matrix,arrows,decorations.pathmorphing}
\usepgfmodule{shapes}

\usetikzlibrary{arrows,decorations.markings}
\usepackage{color}
\usepackage{pgf}
\usepackage{pgffor}

\usepackage{epic}
\usepackage{hyperref}
\usepackage{bbold}

\colorlet{shadecolor}{blue!15}

\newcommand{\beqa}{\begin{eqnarray}}
\newcommand{\eeqa}{\end{eqnarray}}

\def\Tr{\rm Tr}

\newcommand{\beq}{\begin{equation}}
\newcommand{\eeq}{\end{equation}}
\newcommand{\bea}{\begin{eqnarray}}
\newcommand{\eea}{\end{eqnarray}}

\newcommand{\CN}{{\mathcal N}}

\newcommand{\be}{\begin{equation}}
\newcommand{\ee}{\end{equation}}

\newcommand{\bpic}{\begin{tikzpicture}}
\newcommand{\epic}{\end{tikzpicture}}

\begin{document}

\thispagestyle{empty}

\begin{center}
{\Huge Infrared enhancement of supersymmetry in four dimensions} 
\\[15mm]
{\Large{Simone Giacomelli}  
} 
\vskip 6mm
 
\bigskip
{\it  
International Center for Theoretical Physics, Strada Costiera 11, 34151 Trieste, Italy.
\\
INFN, Sezione di Trieste,
Via Valerio 2, 34127 Trieste, Italy 
\\
  }
\vskip 6 mm
e-mail: sgiacome@ictp.it
\bigskip
\bigskip

{\large{\bf Abstract}}\\[5mm]
{\parbox{14cm}{\hspace{5mm}

We study a recently-found class of RG flows in four dimensions exhibiting enhancement of supersymmetry in the infrared, which provides a lagrangian description of several strongly-coupled N=2 SCFTs. The procedure involves starting from a N=2 SCFT, coupling a chiral multiplet in the adjoint representation of the global symmetry to the moment map of the SCFT and turning on a nilpotent expectation value for this chiral. In this note we show that, combining considerations based on 't Hooft anomaly matching and basic results about the N=2 superconformal algebra, it is possible to understand in detail the mechanism underlying this phenomenon and formulate a simple criterion for supersymmetry enhancement which allows us to bypass the analysis with a-maximization. As a byproduct, we propose an algorithm to identify a lagrangian UV completion of a given N=2 SCFT under an RG flow of this type, provided there is one.
}
}
\end{center}
\newpage
\pagenumbering{arabic}
\setcounter{footnote}{0}
\renewcommand{\thefootnote}{\arabic{footnote}}

\tableofcontents

\section{Introduction} 

Recently there has been considerable interest in RG flows which display enhancement of supersymmetry in the infrared in various dimensions. The main purposes of this investigation are, among others, to construct new examples of theories with enhanced supersymmetry, whose dynamics is highly constrained, and provide a new perspective on strongly-coupled models we know already. 

A particularly interesting example of the latter, which constitutes the main focus of this note, is the class of RG flows in four dimensions found recently by Maruyoshi and Song in  \cite{Maruyoshi:2016tqk, Maruyoshi:2016aim}, which provides a lagrangian $\CN=1$ description of a large class of strongly-coupled $\CN=2$ theories called Argyres-Douglas models \cite{Argyres:1995jj}-\cite{Xie:2012hs} (see also \cite{Grover:2013rc}-\cite{Buican:2018ddk} for other examples of RG flows with supersymmetry enhancement in three and four dimensions). The lagrangian description is of course very helpful in getting further insight about the moduli space of the theory and to compute its partition function on various backgrounds.  

Let's discuss more in detail the result of \cite{Maruyoshi:2016tqk, Maruyoshi:2016aim}: starting from a $\mathcal{N}=2$ superconformal theory, we can deform it by adding a chiral multiplet (coupled to the moment map) in the adjoint representation of the flavor symmetry and then giving a nilpotent vev to it. In several cases it turns out that the infrared fixed point has eight supercharges, despite the fact that the new interaction term manifestly breaks extended supersymmetry. Further examples of enhancement of this type have been discussed in \cite{Agarwal:2016pjo}-\cite{Benvenuti:2017bpg} (see also \cite{Benvenuti:2017lle, Benvenuti:2017kud} for the reduction to 3d).

 The main tool discussed in the above references to study these RG flows is a-maximization \cite{Intriligator:2003jj}, which allows us to identify the infrared R-symmetry. From this result one can determine the scaling dimensions of chiral operators and the a, c central charges of the IR fixed point and sometimes, for specific choices of the UV SCFT and nilpotent vev, one recognizes the spectrum and central charges of known $\CN=2$ SCFT's therefore providing good evidence for enhancement of supersymmetry. However, in general one ends up in the IR with a SCFT with irrational scaling dimensions leading to the conclusion that supersymmetry does not enhance. From the a-maximization analysis alone, it is not obvious how to characterize UV SCFTs which exhibit enhancement in the infrared (and for which choices of nilpotent vev) and it is unclear whether supersymmetry enhances or not whenever the scaling dimensions of chiral operators end up being rational but cannot be matched with those of known $\CN=2$ SCFTs. One possible approach to this problem is to study the superconformal index of the theory as in \cite{Evtikhiev:2017heo}: supersymmetry enhancement can occur only if the index has specific properties and one can check whether the required constraints are satisfied or not. However, this approach requires working out the details of the index which quickly becomes computationally challenging as we increase the rank (dimension of the Coulomb Branch) of the UV SCFT and makes it hard to extract a precise pattern, which at the moment is still missing. 

As the above discussion shows, it would be desirable to have a simple criterion to establish (or rule out) supersymmetry enhancement and this is precisely the purpose of the present note.  As we will see, there is indeed a pattern and it can be understood. Our main observation is that, instead of restricting to the a-maximization analysis, one can push further 't Hooft anomaly matching finding several constraints for supersymmetry enhancement which are entirely not obvious from a-maximization alone. Indeed the constraints we find are necessary, but we conjecture our criterion is also sufficient and in the rest of this note we will provide supporting evidence for our claim. We now summarize our approach and the main results. 

\subsection{Strategy and statement of the results} 

In order to illustrate our results, we need to explain some more details about these RG flows: the chiral ring of the theory in the UV includes Higgs Branch (HB) and Coulomb Branch (CB) operators of the UV $\CN=2$ SCFT and some singlets (the components of the adjoint chiral which do not decouple due to spontaneous breaking of the global symmetry). In the IR some singlets and some CB operators hit the unitarity bound and decouple. Whenever supersymmetry enhances, all the CB operators and singlets which remain above the unitarity bound become the CB operators of the IR $\CN=2$ SCFT. This fact holds true for all known cases of susy enhancement and in the present paper we will assume this is always the case. Roughly speaking, we are making an a priori guess about the UV origin of CB operators of the IR SCFT. This assumption is crucial for our construction. 

Using the assumption and imposing supersymmetry enhancement at long distances, we derive in Section \ref{anomalysec}  five nontrivial equations from anomaly matching considerations. One directly tells us that the dimension of the Coulomb Branch is preserved under RG flows of this type. Exploiting then some basic considerations about $\CN=2$ superconformal multiplets (see Section \ref{sec4}), which basically determine the scaling dimensions of chiral operators in the IR, we find that the other four equations derived from anomaly matching can be interpreted as follows: two of them are used to fix the central charges of the IR SCFT\footnote{We would like to point out that our formula for the a,c central charges (\ref{final}) at the IR fixed point are derived under the assumption of supersymmetry enhancement and therefore should be trusted only in this case. Whenever supersymmetry does not enhance in the IR, the outcome of a-maximization will differ from  (\ref{final}).}, one equations enforces the Shapere-Tachikawa relation \cite{Shapere:2008zf} (see also \cite{Argyres:2007tq}) both for the UV and IR $\CN=2$ SCFT's:
\be\label{cbformula}8a-4c=\sum_i(2D_i-1),\ee 
where the sum runs over CB operators and $D_i$ denotes their scaling dimension\footnote{This means that whenever the UV SCFT does not satisfy (\ref{cbformula}) supersymmetry does not enhance in the infrared and analogously, all the theories which violate  (\ref{cbformula}) cannot be realized as IR fixed points of an RG flow of the type we are discussing.}. The last equation is interpreted as a further constraint on the RG flow and in the special case of principal nilpotent vev it can be seen as a characterization of $\CN=2$ SCFT's which exhibit enhancement of supersymmetry in the infrared under this type of flow. The equation reads 
\be 6c-r=\frac{3\beta_G I_{\rho}^p-6(h-D_{max})\sum_{i}(D_i-1)^2}{(h-D_{max})(h+2-D_{max})},\ee 
where $r$ is the rank of the SCFT, $\beta_G$ is half the flavor central charge, $h$ is the Coxeter number of the global symmetry group, $I_{\rho}^p$ is the embedding index of the principal nilpotent orbit (see Table \ref{embindx}) and $D_{max}$ is the dimension of the CB operator of largest dimension.  It would be interesting to find a simple interpretation of this equation. 

Combining all these constraints we formulate a criterion for supersymmetry enhancement in Section \ref{sec4}. All the RG flows for which enhancement of supersymmetry has been excluded in \cite{Evtikhiev:2017heo} using superconformal index arguments can be easily ruled out using our criterion. In Appendix \ref{maxcheck} we prove that whenever our criterion is satisfied, the a and c central charges predicted by our construction reproduce the a-maximization analysis. This in particular tells us that the proposed criterion is strictly more informative than a-maximization. Finally, using (\ref{cbformula}) and our anomaly matching equations,  in Section \ref{sec4} we prove that all RG flows which exhibit enhanced supersymmetry in the IR satisfy the following simple relation: 
$$(6c'-r)(4a-5c)=(6c-r)(4a'-5c')$$
where $r$ is the dimension of the Coulomb Branch (which is the same in the UV and IR), a and c are the central charges of the UV SCFT and a', c' those of the IR SCFT. The remarkable feature of this equation is that it does not depend on the global symmetry of the UV theory, although the definition of the RG flow exploits this information. In Appendix \ref{checklit} we check that this formula is satisfied by all known RG flows with enhanced supersymmetry using the formalism of \cite{Giacomelli:2017ckh}. This equation can be exploited to formulate a criterion for the existence of a UV lagrangian completion of a given $\CN=2$ SCFT under an RG flow of this type. We explain this in Section \ref{ciaociao}.

\section{Anomaly matching}\label{anomalysec}

\subsection{R-symmetry of the infrared SCFT}\label{sec2}

We start by recalling how to determine the R-symmetry at the infrared fixed point for the class of RG flows of interest. First of all we exploit the fact that every $\CN=2$ superconformal theory has two canonical $U(1)$ global symmetries (the $U(1)_R$ group $R_{\CN=2}$ and the cartan of the $SU(2)_R$ symmetry $I_3$). If the theory has a further global symmetry group $G_F$, we can add a chiral multiplet $M$ transforming in the adjoint of $G_F$ and turn on the superpotential term 
\be\label{supp}\mathcal{W}=\int d^2\theta \Tr(M\mu_{G}),\ee
where $\mu_G$ is the moment map associated with the symmetry $G_F$. From the interaction (\ref{supp}) we conclude that $M$ is not charged under $I_3$ and has charge 2 under $R_{\CN=2}$. If we now give a nilpotent vev to $M$, we also need to consider the Cartan $\rho(\sigma_3)$ of the $SU(2)$ embedding labelling the nilpotent orbit (our convention is $\langle M\rangle=\rho(\sigma^+)$). Out of these three $U(1)$'s, $I_3$ and the combination $R_{\CN=2}-2\rho(\sigma_3)$ are not broken by the vev and we assume the $U(1)$ R-symmetry of the IR fixed point is a combination of them. We can parametrize it as follows: 
\be\label{rtrial}R_{\epsilon}=\frac{1+\epsilon}{2}(R_{\CN=2}-2\rho(\sigma_3))+(1-\epsilon)I_3.\ee 
The value of $\epsilon$ for which (\ref{rtrial}) is identified with the infrared R-symmetry can be determined via a-maximization and will be denoted as $\epsilon_*$ from now on. 
When we turn on the vev, some components of $M$ become the Goldstone multiplets associated with the spontaneous symmetry breaking and those which remain coupled to the theory can be described as follows (see \cite{Gadde:2013fma, Agarwal:2014rua, Agarwal:2013uga} for a derivation): the adjoint representation of the global symmetry $G_F$ decomposes as the direct sum of irreducible representations of the $SU(2)$ subgroup which labels the nilpotent orbit and all the states of the $SU(2)$ irreps become Goldstone multiplets except the lowest weight states. We therefore have a one-to-one correspondence between $SU(2)$ representations and singlets and from now on we will denote the singlets with the spin $s$ of the corresponding $SU(2)$ irreducible representation. 

A crucial observation for us is that the CB operators of the UV SCFT and the singlets coming from the adjoint chiral $M$ are uncharged under $I_3$. Some of these operators hit the unitarity bound in the infrared and decouple, whereas all the others become (if supersymmetry enhances) CB operators of the IR SCFT. The charge under (\ref{rtrial}) of the UV CB operators is $(1+\epsilon)D$ (where $D$ is the scaling dimension in the UV) and the singlets have charge $(1+\epsilon)(s+1)$ (where again $s$ is the spin of the corresponding $SU(2)$ representation). Notice that all these operators have charge of the form $(1+\epsilon)k$, where $k$ is half the charge of the operator under $R_{\CN=2}-2\rho(\sigma_3)$. Once the UV SCFT and the nilpotent orbit are specified, the set of allowed values of $k$ is fixed and can be considered part of the defining data of our problem. In general we have: 
\be\label{kpara}k\equiv\;\left\{\begin{array}{l} 
   D\;\; (\text{for UV CB operators}) \\
  \\
  s+1\;\; (\text{for singlets})\\
\end{array}\right. \ee
where indeed $D$ is the scaling dimension of the CB operator and $s$ the spin of the representation labelling the singlet. 
The operators which violate the unitarity bound and decouple are those with $k<k_c$, where 
\be\label{kcrit}k_c=\frac{2}{3+3\epsilon_*}.\ee 
Our goal is to find, without using any extremization principle, a general formula for the value of $\epsilon_*$, under the assumption that supersymmetry enhances in the IR. Concretely we are going to assume that the cartan generators of the infrared $U(2)$ R-symmetry are linear combinations of the two unbroken $U(1)$'s $R_{\CN=2}-2\rho(\sigma_3)$ and $I_3$.

We will make use of the following well-known formulas for $\CN=2$ SCFT's \cite{Shapere:2008zf, Kuzenko:1999pi}
\be\label{anomn2} \Tr R_{\CN=2}^3=\Tr R_{\CN=2}=48(a-c);\quad \Tr R_{\CN=2}I_3^2=4a-2c.\ee 
Other 't Hooft anomalies are trivial, in particular the triviality of $\Tr I_3^3$ will be crucial for our argument.

Imposing supersymmetry enhances in the IR, we know that (\ref{rtrial}) is the subgroup of $U(1)_R\times SU(2)_R$ such that the scaling dimension of all chiral primary operators in the IR SCFT reads $D(O)=\frac{3}{2}R_{\epsilon_*}(O)$. This is the R-symmetry of the $\CN=1$ subalgebra which is manifestly preserved along the flow and in terms of the generators $R'_{\CN=2}$ and $I'_3$ of the $U(1)\times SU(2)$ R-symmetry of the IR SCFT it reads 
\be\label{rsym1}R_{\epsilon_*}=\frac{1}{3}R'_{\CN=2}+\frac{4}{3}I'_3.\ee 
The other cartan generator of $U(1)_R\times SU(2)_R$ R-symmetry, which we may write in the form 
\be\label{testu1}R'_{\CN=2}-2I'_3,\ee 
is (from the perspective of the $\CN=1$ subalgebra preserved by the flow) just a global $U(1)$ symmetry. This means in particular that all the components of the $\CN=1$ multiplets have the same charge under it.
This fact can be exploited to write this $U(1)$ generator in terms of $R_{\CN=2}-2\rho(\sigma_3)$ and $I_3$: 
by acting with the manifest supercharge on the lowest component of a chiral multiplet with charge $(r,\tilde{r})$ under $R_{\CN=2}-2\rho(\sigma_3)$ and $I_3$ respectively, we get another component of the multiplet whose charges are $(r-1,\tilde{r}-1/2)$. 
Requiring now the two components to have the same charge, we conclude that the generator we are after can be written as follows: 
\be\label{trialu1}R'_{\CN=2}-2I'_3=\alpha(2I_3-R_{\CN=2}+2\rho(\sigma_3)).\ee 
In order to fix $\alpha$ it suffices to notice that the charge of CB operators of the IR SCFT under (\ref{testu1}) is three times that under (\ref{rsym1}). Exploiting now the fact that IR CB operators have charge of the form $(1+\epsilon)k$ under (\ref{rtrial}), we conclude that $\alpha=-\frac{3+3\epsilon_*}{2}$. 
Using now (\ref{rsym1}) and (\ref{trialu1}), we can extract a formula relating the UV and IR generators of the $U(1)_R\times SU(2)_R$ symmetry: 
\be\label{rir1}R'_{\CN=2}=\frac{3}{2}(1+\epsilon_*)(R_{\CN=2}-2\rho(\sigma_3))-(1+3\epsilon_*)I_3;\quad I'_3=I_3.\ee 
We therefore conclude that $I_3$ becomes the cartan of the $SU(2)_R$ symmetry in the infrared; in particular $\Tr I_3^3$ is trivial in the infrared SCFT. Equation (\ref{rir1}) constitutes our main tool for the anomaly matching analysis which follows.  

\subsection{Preservation of the rank and its consequences}\label{sec31}

Based on the results derived so far, we can immediately conclude that the 't Hooft anomaly $\Tr I_3^3$ receives a nontrivial contribution in the UV from the singlets only and, under the assumption of supersymmetry enhancement at long distances, only from decoupled operators in the IR. Since these are free fields their contribution to the anomaly is just the cube of the charge of the fermionic components under $I_3$, and this is equal to $-1/2$ both for the singlets and the decoupled operators. By anomaly matching we therefore conclude that 

\begin{shaded} 
\noindent The number of singlets we add in the UV is equal to the number of operators which decouple in the IR, hence the rank of the IR SCFT (the number of Coulomb Branch generators) is the same as the rank of the UV SCFT. 
\end{shaded}

We refer to this property as ``rank condition''. We stress that this conclusion holds only when supersymmetry enhances in the infrared; if this is not so, there is no reason to impose that the IR SCFT does not contribute to the 't Hooft anomaly $\Tr I_3^3$. 

This observation alone is enough to gain some nontrivial insight about the pattern of SUSY enhancement observed in the literature. Just to illustrate how this works, let us consider the linear superconformal quivers: 
$$\boxed{m}-SU(N+m)-SU(2N+m)-\boxed{3N+m}$$
It was observed that by turning on a principal nilpotent vev for the $SU(3N+m)$ global symmetry we end up with an $\CN=2$ SCFT (plus decoupled free fields) in the IR only for $m=0,1$. The case $m>1$ instead does not exhibit enhancement. The rank condition we have just derived allows us to understand immediately what goes wrong in the latter case. For simplicity we focus on the case $N=2$, the generalization to other values of $N$ is straightforward and works exactly in the same way. 

For $m=0$, $N=2$ the theory has rank 4 and the corresponding values of $k$ are $2,2,3,4$ (see (\ref{kpara})). Choosing the principal nilpotent orbit for $SU(6)$ we get 5 singlets with $k=2,3,4,5,6$. If the result of a-maximization is such that $3\leq k_c<4$, then 5 operators violate the unitarity bound and decouple in the IR. In this case the rank condition is satisified and SUSY enhancement is not ruled out. This is indeed the outcome of the a-maximization analysis. 
Similarly, for $m=1$, $N=2$ the CB operators have $k=2,3,2,3,4,5$ and the singlets $k=2,3,4,5,6,7$ and again the rank condition is satisfied if $3\leq k_c<4$. 

On the other hand, for $m=N=2$ the CB operators have $k=2,3,4,2,3,4,5,6$ and the 7 singlets $k=2,3,4,5,6,7,8$. In this case, no matter what the outcome of a-maximization is, it is impossible to satisfy the rank condition which requires 7 operators to decouple. Indeed, depending on the value of $k_c$, we find that the number of decoupled operators can only be 0,3,6 or at least 9. In this sense the violation of the rank condition, which was derived under the assumption of infrared enhancement, immediately tells us that supersymmetry will not enhance in this case. Models with higher values of $m$ and $N$ and the case of longer quivers can be treated along the same line. 

Another interesting observation (already mentioned in \cite{Giacomelli:2017ckh}) is that this argument neatly explains why SQCD with gauge group $SO(N)$ does not exhibit enhancement\footnote{This fact was observed via a-maximization in \cite{Agarwal:2016pjo, Agarwal:2017roi}. Our goal is to provide a rationale for this observation.}: in the case of e.g. $SO(5)$ SQCD with 3 hypers in the vector representation the CB operators have $k=2,4$. If we turn on a principal nilpotent vev for the $USp(6)$ global symmetry the singlets have $k=2,4,6$ and again the rank condition cannot be satisfied: the number of decoupled operators has to be either 0,2 or at least 4; in any case not 3. 

Finally, we can easily see that Minahan-Nemeschansky theories \cite{Minahan:1996fg, Minahan:1996cj} cannot have a UV lagrangian completion, at least with an RG flow of this type: the rank condition implies the UV theory should have rank one but the only lagrangian theories with this property are $SU(2)$ SQCD with four flavors or $SU(2)$ $\CN=4$ SYM and neither of these models can flow to Minahan-Nemeschansky $E_n$  theories. 

Notice that the rank condition is a necessary but clearly not sufficient condition. For example, if we consider the linear quiver discussed above with $m=3$ and $N=2$ and consider the next-to-maximal nilpotent vev for the $SU(9)$ global symmetry, we find that the rank condition is satisfied provided $4\leq k_c<\frac{9}{2}$, so supersymmetry enhancement cannot be ruled out by this consideration alone. On the other hand, by direct inspection one can check there is no enhancement in this case. We therefore need a more refined criterion and we will explain in the rest of this note how anomaly matching can be used to derive it.

\subsection{Other 't Hooft anomalies}

We can get more quantitative information about the infrared $\CN=2$ SCFT by considering the matching of $\Tr R'_{\CN=2}$ and $\Tr R'_{\CN=2}I_3^2$. This time the SCFTs in the UV and IR contribute and using (\ref{anomn2}),(\ref{rir1}) (we denote with a and c the central charges in the UV and with $a'$ and $c'$ those in the IR) we find the equations 
\be\label{anmatch1}24(1+\epsilon_*)(a-c)+(1+\epsilon_*)\sum_i(s_i+1)=16(a'-c')+(1+\epsilon_*)\sum_{j,dec}k_j,\ee 
\be\label{anmatch2}(1+\epsilon_*)(2a-c)+\frac{1+\epsilon_*}{4}\sum_i(s_i+1)=\frac{4a'-2c'}{3}+\frac{1+\epsilon_*}{4}\sum_{j,dec}k_j.\ee 
The first relation comes from $\Tr R'_{\CN=2}$ and the second from $\Tr R'_{\CN=2}I_3^2$. In the above formulas $\sum_i$ denotes the sum over singlets and $s_i$ is the spin of $SU(2)$ representations, whereas $\sum_{j,dec}k_j$ denotes the sum over decoupled operators of the corresponding $k_i$'s. Notice that in deriving (\ref{anmatch1}) and (\ref{anmatch2}) we made use of the rank condition, which implies that the number of singlets is equal to the number of decoupled operators. This observation is used to remove a constant term from both sums. Combining these two equations we find the relation 
\be\label{anmatch}\frac{3}{2}(1+\epsilon_*)(4a-5c)=4a'-5c'.\ee 
The anomaly matching condition for $\Tr(R'_{\CN=2})^2I_3$ can be written in the form 
\be\label{anr2}3(1+\epsilon_*)\left(8a-4c+\sum_i(s_i+1)^2-\sum_{j,dec}k_j^2\right)-2\left(8a-4c+\sum_i(s_i+1)-\sum_{j,dec}k_j\right)=0.\ee 
The equation for $\Tr(R'_{\CN=2})^3$ is somewhat complicated and we find it more convenient to combine it with (\ref{anmatch1}) to get an expression which does not depend on the central charges $a'$, $c'$ of the IR SCFT. The resulting equation for $\Tr(R'_{\CN=2})^3-\Tr R'_{\CN=2}$ reads: 
\be\label{anr3}\begin{array}{ll} 
   0= & 27(1+\epsilon_*)^2\left(12a-9c-\frac{3}{2}\beta_G I_{\rho}+\sum_i(s_i+1)^3-\sum_{j,dec}k_j^3\right)+ 6\left(6c+\sum_i(s_i+1)-\sum_{j,dec}k_j\right)  \\
   & -27(1+\epsilon_*)\left(8a-4c+\sum_i(s_i+1)^2-\sum_{j,dec}k_j^2\right)\\
   \end{array}\ee
and using (\ref{anr2}) we can bring it to the simpler form 
\be\label{an3final}12a-9c+\sum_i(s_i+1)-\sum_{j,dec}k_j=\frac{(3+3\epsilon_*)^2}{4}\left(12a-9c-\frac{3}{2}\beta_G I_{\rho}+\sum_i(s_i+1)^3-\sum_{j,dec}k_j^3\right).\ee
In the last two equations $I_{\rho}$ denotes the embedding index of the $U(1)$ subgroup $\rho(\sigma_3)$ inside $G_F$ and $\beta_G$ is the $G_F$ flavor central charge divided by two (or equivalently the contribution of the SCFT to the beta function if we gauge $G_F$). These quantities appear because 
$$\Tr R_{\CN=2}\rho(\sigma_3)^2=I_{\rho}\Tr R_{\CN=2}G_F^2=-\beta_G I_{\rho}.$$ 
The equations are written in this form for later convenience.

\section{Criterion for enhancement}\label{sec4} 

In order to gain further insight, we combine the 't Hooft anomaly matching of the previous section with our knowledge about the structure of $\CN=2$ multiplets. This will allow us to understand in detail the mechanism underlying supersymmetry enhancement in the infrared. 

\subsection{Analysis of short multiplets}
 
We start by observing that all the CB operators of the UV theory fit into $\mathcal{E}_{r(0,0)}$ multiplets (see \cite{Dolan:2002zh}) which can be decomposed into chiral multiplets of the $\CN=1$ subalgebra manifestly preserved by the flow (we use the same conventions as in \cite{Maruyoshi:2016aim}). The CB operator itself is the lowest component of the chiral multiplet\footnote{We adopt the standard notation (see \cite{Dolan:2002zh, Gadde:2010en} for details) where $\mathcal{B}_{r(j_1,j_2)}$ denotes the chiral multiplet with spin $(j_1,j_2)$ and charge $r$ under the $U(1)$ R-symmetry of the $\CN=1$ superconformal algebra. In the $\CN=2$ case at hand this is the combination $\frac{1}{3}R_{\CN=2}+\frac{4}{3}I_3$.} $\mathcal{B}_{\frac{r}{3}(0,0)}$. This is the chiral multiplet uncharged under $I_3$ which may decouple in the infrared. The $\mathcal{E}$ multiplet contains other three chirals: $\mathcal{B}_{\frac{r+1}{3}(0,\pm\frac{1}{2})}$ with charge $\frac{1}{2}$ under $I_3$ and $r-1$ under $R_{\CN=2}$, and  
a multiplet $\mathcal{B}_{\frac{r+2}{3}(0,0)}$ with charge $1$ under $I_3$ and $r-2$ under $R_{\CN=2}$. 

If supersymmetry enhances in the infrared, necessarily the multiplets $\mathcal{B}_{\frac{r+1}{3}(0,\pm\frac{1}{2})}$ and $\mathcal{B}_{\frac{r+2}{3}(0,0)}$ have to recombine in the IR with a chiral multiplet (either a singlet or a UV CB operator) to rebuild the $\CN=2$ superconformal multiplet. Moreover, extended supersymmetry also requires the charge of this chiral under $R'_{\CN=2}$ in (\ref{rir1}) (which is the $U(1)_R$ symmetry at the infrared fixed point) to be equal to that of $\mathcal{B}_{\frac{r+1}{3}(0,\pm\frac{1}{2})}$ plus one. We therefore find the equation 
$$ 3(1+\epsilon_*)k'=\frac{3}{2}(1+\epsilon_*)(2k-1)-\frac{1+3\epsilon_*}{2}+1,$$ 
where $k$ is the dimension of the UV CB operator and $k'$ the corresponding parameter for the chiral which recombines in the infrared with the $\mathcal{B}$ multiplets (see (\ref{kpara})). We therefore conclude that 
\be\label{1to1}k'-k=-\frac{1+3\epsilon_*}{3+3\epsilon_*}.\ee 
Notice that the r.h.s. is positive if $-1<\epsilon_*\leq-\frac{1}{3}$ and this inequality is implied by the rank condition: if the upper bound is violated none of the singlets and CB operators decouple in the IR and if the lower bound is violated all of them decouple. 

As we have explained in the previous section, the rank condition is the statement that there is a one-to-one correspondence between UV and IR CB generators; the above reasoning exhibits explicitly a bijection between the two sets: the map sending the operator labelled by $k$ to that labelled by $k'$ (notice that from the obvious inequality $k>1$, (\ref{1to1}) tells us that the operator $k'$, whose dimension in the IR is $\frac{3}{2}(1+\epsilon_*)k'$, does not violate the unitarity bound). This map is particularly convenient because the difference between $k$ and the corresponding $k'$ given by (\ref{1to1}) does not depend on the specific UV CB operator we are considering. This in particular implies the following observation: if the UV theory has $n$ operators of a given dimension $d$, the IR theory will necessarily include $n$ operators with dimension 
$$\frac{3}{2}(1+\epsilon_*)\left(d-\frac{1+3\epsilon_*}{3+3\epsilon_*}\right).$$ 
Said differently, the degeneration of the CB spectrum is preserved under an RG flow of this type. This will prove useful in Section \ref{ciaociao} when we discuss UV lagrangian completions of a SCFT. 

The above discussion has an immediate payoff: consider the UV CB operator of largest dimension, whose dimension we denote as $D_{max}$. The corresponding IR CB operator under the map described above will become the CB operator in the IR with largest scaling dimension because of (\ref{1to1}). We now want to argue that, unless the RG flow is trivial, the CB operator with largest dimension in the IR arises from a singlet in the UV: suppose this is not the case, then it has to be a CB operator in the UV as well. However, the operator with dimension $D_{max}$ has the largest dimension by definition and consequently has the largest possible value of $k$ among UV CB operators. But then $k'_{max}-k_{max}$ cannot be strictly positive and from (\ref{1to1}) we conclude that 
$$k'=k;\quad \epsilon_*=-\frac{1}{3},$$ 
therefore all operators retain the dimension they have in the UV, all the singlets decouple and from (\ref{anmatch1}), (\ref{anmatch2}) we conclude $a'=a$ and $c'=c$. This is just the trivial RG flow as expected. In order to avoid this conclusion we need $k'_{max}-k_{max}$ to be strictly positive, but this can be the case only if there is a singlet with $k$ larger than $k_{max}$, which implies 
\be\label{singc} s_{max}+1>D_{max},\ee 
where $s_{max}$ is the largest spin appearing in the decomposition of the adjoint of $G_F$ in terms of $SU(2)$ representations. We therefore conclude that 
\begin{shaded} 
\noindent Supersymmetry enhancement in the infrared requires the presence of at least a singlet with charge under $R_{\CN=2}-2\rho(\sigma_3)$ strictly larger than all Coulomb Branch operators. 
\end{shaded}
Before proceeding, let us pause to discuss a nontrivial application of this property: consider the quiver 
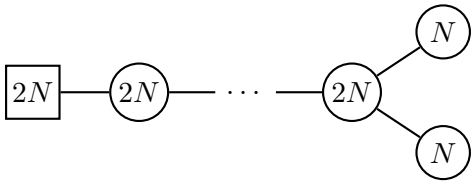
\begin{center}
\begin{tikzpicture}[->,thick, scale=0.4]
\node[rectangle, draw, minimum height=20pt, minimum width=20pt, inner sep=1.7](L1) at (-10.5,0){ $2N$};
\node[circle, draw, inner sep=1.7](L2) at (-7,0){ $2N$};
\node[](L3) at (-3.5,0){\dots};
\node[circle, draw, inner sep=1.7](L4) at (0,0){ $2N$};
\node[circle, draw, minimum height=20pt, inner sep=1.7](L5) at (3,2){$N$};
\node[circle, draw, minimum height=20pt, inner sep=1.7](L6) at (3,-2){$N$};

\draw[-] (L1) -- (L2);
\draw[-] (L3) -- (L2);
\draw[-] (L3) -- (L4);
\draw[-] (L4) -- (L5);
\draw[-] (L4) -- (L6);
\end{tikzpicture}
\end{center}
where circles with $N$ inside denote $SU(N)$ gauge groups and edges represent bifundamental hypermultiplets. If we turn on a nilpotent vev for the $SU(2N)$ global symmetry rotating the flavors at the left end of the quiver, we necessarily have the relation $s_{max}+1\leq 2N$ and the inequality is saturated only for the principal nilpotent vev. Since in this quiver there are CB operators of dimension $2N$, the inequality (\ref{singc}) is violated and therefore we conclude that supersymmetry will not enhance in the infrared. This conclusion holds for every value of $N$ and every choice of nilpotent vev.

\subsection{Formulation of the criterion}

Since we have just argued that all nontrivial cases of supersymmetry enhancement occur when (\ref{singc}) is satisfied, we concentrate on this case from now on. From (\ref{1to1}) we can directly predict the value of $\epsilon_*$: 
\be\label{amaxf}\frac{3}{2}(1+\epsilon_*)=\frac{1}{2+s_{max}-D_{max}}.\ee 
From (\ref{anmatch1}) and (\ref{anmatch2}) we then find that 
\be\label{final} a'=\frac{24a+5r(s_{max}+1-D_{max})}{24(2+s_{max}-D_{max})};\quad c'=\frac{6c+r(s_{max}+1-D_{max})}{6(2+s_{max}-D_{max})}.\ee 
The conclusion is that we have achieved, under the assumption of enhancement of supersymmetry in the infrared, a formula expressing $\epsilon_*$ (and also the central charges of the IR SCFT) in terms of the defining data of the RG flow (data of the UV SCFT and the choice of nilpotent vev, which fixes the set of spins $s_i$). 

In deriving (\ref{final}) we have used (\ref{1to1}) to rewrite $\sum_i(s_i+1)-\sum_{j,dec}k_j$. The idea is the following: the set of singlets and UV CB operators can also be interpreted as the set of IR CB operators and decoupled fields, therefore we have the relation 
\be\label{reorder}\sum_{l,UV}k_l^n+\sum_i(s_i+1)^n=\sum_{l,IR}(k'_l)^n+\sum_{j,dec}k_j^n,\ee 
for every $n$. $\sum_{l,UV}$ and $\sum_{l,IR}$ denote indeed the sum over UV and IR CB operators respectively. Setting $n=1$ we see that 
\be\label{idesimp} \sum_i(s_i+1)-\sum_{j,dec}k_j=\sum_{l,IR}k'_l-\sum_{l,UV}k_l=\sum_{l,UV}(k'_l-k_l)=r(s_{max}+1-D_{max}).\ee 
In the second equality we exploited the one-to-one correspondence between CB operators in the UV and in the IR to rewrite everything as a single sum and in the last equality we exploited (\ref{1to1}), which guarantees that all the terms we are summing over are equal to $s_{max}+1-D_{max}$. Analogously, we can exploit (\ref{reorder}) to simplify (\ref{anr2}) and (\ref{an3final}). For instance, we can write 
\be\label{cnuova} \sum_i(s_i+1)^2-\sum_{j,dec}k_j^2=\sum_{l,UV}((k'_l)^2-k_l^2)=-\frac{1+3\epsilon_*}{3+3\epsilon_*}\sum_{l,UV}\left(2k_l-\frac{1+3\epsilon_*}{3+3\epsilon_*}\right).\ee 
We can now notice that for UV CB operators the value of $k$ coincides with the scaling dimension in the UV, therefore the above expression is proportional to the sum over CB operators of their scaling dimension. We rewrite it as follows: 
\be\label{delta}\sum_{l,UV}2k_l=8a-4c+r-2\Delta,\ee 
where $\Delta$ ``parametrizes the deviation'' from the Shapere-Tachikawa relation (\ref{cbformula})
$$8a-4c=\sum_i(2D_i-1).$$
Plugging this in (\ref{cnuova}) and then substituting in (\ref{anr2}) we find 
\be\frac{3+3\epsilon_*}{2}\left(8a-4c-\frac{1+3\epsilon_*}{3+3\epsilon_*}(8a-4c-2\Delta)\right)=8a-4c,\ee 
which clearly reduces to $$(1+3\epsilon_*)\Delta=0.$$ 
This equation is telling us that the UV theory satisfies the Shapere-Tachikawa relation, unless $\epsilon_*=-1/3$. As we have seen, this value corresponds to the trivial RG flow in which UV and IR SCFT's coincide. This is indeed expected: if the numerical prefactor did not vanish in the case of the trivial RG flow, we could use this equation to conclude that the Shapere-Tachikawa relation holds for all $\CN=2$ SCFT's and this is known to be wrong (see \cite{Aharony:2016kai, Argyres:2016yzz}). 

Analogously, we can prove that the deviation $\Delta'$ from the Shapere-Tachikawa formula for the IR SCFT vanishes: we first rewrite (\ref{anr2}) in terms of the central charges of the IR SCFT using (\ref{anmatch2}) and then rewrite (\ref{cnuova}) as follows: 
\be\label{cnuova2}\sum_i(s_i+1)^2-\sum_{j,dec}k_j^2=-\frac{1+3\epsilon_*}{3+3\epsilon_*}\sum_{l,IR}\left(2k'_l+\frac{1+3\epsilon_*}{3+3\epsilon_*}\right).\ee 
Observing now that 
\be \frac{3}{2}(1+\epsilon_*)\sum_{l,IR}2k'_l=\sum_{l,IR}2D'_l=8a'-4c'+r-2\Delta'\ee
where $D'_l$ denotes the scaling dimension of CB operators in the IR SCFT, we find that (\ref{anr2}) reduces to 
$$(1+3\epsilon_*)\Delta'=0.$$
In conclusion, we have learned that only $\CN=2$ SCFT's which satisfy (\ref{cbformula}) can display supersymmetry enhancement (at least with this type of RG flow) in the IR and the infrared fixed point has the same feature.  

Equation (\ref{an3final}) can be handled similarly: setting $n=3$ in (\ref{reorder}) we can rewrite $\sum_i(s_i+1)^3-\sum_{j,dec}k_j^3$ in terms of the scaling dimensions of CB operators. We get a term involving the sum of their scaling dimension, which can be rewritten as in (\ref{delta}) and another term involving the sum of the square of scaling dimensions of CB operators. After some algebra, we can rewrite (\ref{an3final}) as follows: 
\be\label{an44}(6c-r)\frac{(3\epsilon_*+1)(3\epsilon_*+5)}{(3+3\epsilon_*)^2}=-3\beta_G I_{\rho}-2\frac{3\epsilon_*+1}{1+\epsilon_*}\sum_{i}(D_i-1)^2,\ee 
where $D_i$ are the dimensions of UV CB operators.

If we specialize to the case of principal nilpotent vev, (\ref{an44}) becomes 
\be\label{an441}6c-r=\frac{3\beta_G I_{\rho}^p-6(h-D_{max})\sum_{i}(D_i-1)^2}{(h-D_{max})(h+2-D_{max})}.\ee 
Here we used the fact that $s_{max}+1$ is equal to the Coxeter number $h$ in the case of the principal nilpotent vev. The embedding index for the principal nilpotent orbit $I_{\rho}^p$ is \cite{Embindex}
\be\label{embindx}\begin{array}{|c|c|c|c|c|c|}
\hline 
G_F & ADE & B_n & C_n & G_2 & F_4 \\
\hline 
I_{\rho}^p & \frac{h Dim(G_F)}{6} & \frac{n(n+1)(2n+1)}{3} & \frac{n(4n^2-1)}{3} & 28 & 156 \\
\hline 
\end{array}\ee 
Equation (\ref{an441}) only depends on the data of the UV theory and can therefore be seen as a characterization of $\CN=2$ SCFT's which exhibit enhancement of supersymmetry upon turning on a principal nilpotent vev for their global symmetry $G_F$\footnote{It has been ``experimentally'' observed that when the principal nilpotent vev does not lead to supersymmetry enhancement in the infrared, other choices of nilpotent vev do not work either. This is true in all known cases, although we do not know how to derive this statement.}. 

We now have all the ingredients to formulate our criterion for supersymmetry enhancement. In order to illustrate the underlying idea, let us revisit the model discussed at the end of Section \ref{sec31}: 
$$\boxed{3}-SU(5)-SU(7)-\boxed{9}$$ 
This theory has rank 10 and the CB operators have $k=2,3,4,5,2,3,4,5,6,7$. If we consider the next-to-maximal nilpotent vev labelled by the partition $(8,1)$, we get ten singlets with $k=1,2,3,4,9/2,9/2,5,6,7,8$. Imposing that the rank condition is satisfied we conclude that 10 operators should fall below the unitarity bound and decouple and, as we have claimed in Section \ref{sec31}, this happens if $4\leq k_c<9/2$ or equivalently if 
\be\label{inter}\frac{2}{9}< \frac{3}{2}(1+\epsilon_*)\leq\frac{1}{4}.\ee 
We clearly see that SUSY enhancement in this case cannot be ruled out using the rank condition only. However, since $D_{max}=7$ and $s_{max}=7$, from (\ref{amaxf}) we find 
\be\label{inter2}\frac{3}{2}(1+\epsilon_*)=\frac{1}{2},\ee 
which does not lie in the interval (\ref{inter}). This should be interpreted as follows: the assumption of supersymmetry enhancement in the IR on the one hand tells us that the rank condition should hold and therefore we have the bound (\ref{inter}); on the other it leads to (\ref{inter2}) which is not consistent with the bound. We thus reach a contradiction, meaning that the assumption of enhancement cannot be correct. Coherently with the a-maximization analysis, we conclude that supersymmetry does not enhance in the case at hand. 

We can now formulate our criterion for supersymmetry enhancement: 
\begin{enumerate} 
 \item Given an $\CN=2$ SCFT satisfying (\ref{cbformula}) with global symmetry $G_F$ and a choice of nilpotent vev for the chiral multiplet in the adjoint of $G_F$, compute the value of $\epsilon_*$ using (\ref{amaxf}) and check that the result is compatible with the rank condition. 
 \item Check that the set of operators with $k>k_c$ is compatible with (\ref{1to1}), or in other words check that $k+D_{max}-s_{max}-1$ is always equal to the dimension of a UV CB operator. 
 \item Check that equation (\ref{an44}) is satisfied.
\end{enumerate}
The arguments we have given show that these three requirements are necessary and if the RG flow satisfies all of them, we conjecture that the infrared fixed point is an $\CN=2$ SCFT with $(a,c)$ central charges determined by (\ref{final}). As supporting evidence we will show in Appendix \ref{maxcheck} that (\ref{amaxf}) reproduces the outcome of a-maximization provided the RG flow passes our criterion. 

We would like to remark that there are theories satisfying step 1 but not step 2 and there are also theories satisfying the first two steps but not the third. An example of the former case is the $SO(7)\times USp(8)$ linear quiver 
$$\boxed{SU(2)}-SO(7)-USp(8)-\boxed{SO(13)}$$ 
The global symmetries are indicated in the boxes. If we turn on a principal nilpotent vev for $SO(13)$ we have singlets with $k=2,4,6,8,10,12$ and the CB operators have $k=2,4,6,2,4,6,8$. From (\ref{amaxf}) we find $\frac{3}{2}(1+\epsilon^*)=\frac{1}{5}$ and therefore the theory satisfies step 1. An example of the latter case is given by the quiver 
 \begin{center}
\begin{tikzpicture}[->,thick, scale=0.4]
\node[rectangle, draw, minimum height=20pt, minimum width=20pt](L1) at (4.5,0){ $USp(8)$};
\node[](L4) at (0,0){ $SO(8)$};
\node[](L5) at (-3,2){$SU(2)$};
\node[](L6) at (-3,-2){$SU(2)$};

\draw[-] (L1) -- (L4);
\draw[-] (L4) -- (L5);
\draw[-] (L4) -- (L6);
\end{tikzpicture}
\end{center}
where again we report the global symmetry in the box. If we activate a principal nilpotent vev for $USp(8)$ the singlets have $k=2,4,6,8$ and the CB operators $k=2,2,2,4,4,6$. Both step 1 and step 2 are satisfied but (\ref{an44}) does not hold. In both the above cases we conclude that supersymmetry does not enhance.

Let us now discuss how to apply our criterion in the simple case of rank one theories. As we will now see, implementing the first step is enough to select the RG flows which exhibit supersymmetry enhancement in the infrared. 

\subsection{Testing rank one theories} 

If we start in the UV from a rank 1 SCFT, we know that the IR fixed point should have the same property due to the rank condition  and from (\ref{singc}) we know that the IR CB operator originates from the singlet $s_{max}$ (we are neglecting the trivial RG flow).  

We can now observe that some nilpotent orbits can be ruled out a priori because they are not compatible with the rank condition: sometimes in the decomposition of the adjoint representation of $G_F$ two or more $SU(2)$ representations with highest spin appear. This is for example the case of the nilpotent orbit of $SO(8)$ labelled by the partition $(5,3)$: the highest spin appearing is 3 and there are two representations of that spin. All such nilpotent orbits can be discarded.

In order to discuss the remaining cases, we use (\ref{amaxf}): 
\be\label{cciao}\frac{3}{2}(1+\epsilon_*)=\frac{1}{2+s_{max}-D},\ee 
Notice as a simple consistency check that the singlet (whose dimension in the IR is $\frac{3}{2}(1+\epsilon_*)(s_{max}+1)$) is above the unitarity bound because of the obvious inequality $D>1$ for the dimension of the UV CB operator. 

We should now demand that the UV CB operator and the singlet(s) associated with the second highest spin $s_2$ decouple (in other words we demand that only the singlet $s_{max}$ remains above the unitarity bound). This is the nontrivial step coming from enforcing the rank condition. From (\ref{cciao}) we conclude that the two inequalities read: 
\be\label{nilc}s_1\geq 2D-2;\quad s_2\leq s_1+1-D.\ee 
Roughly speaking, $s_1$ should be large enough and $s_2$ should not be too large. This is an algebraic constraint on the nilpotent orbit and it is easy to check that only the cases which exhibit enhancement of supersymmetry in the infrared are consistent with the above inequalities (compare with \cite{Maruyoshi:2016aim, Agarwal:2016pjo, Giacomelli:2017ckh}): 
\begin{itemize}
\item For $E_8$ MN $D=6$ and only the principal nilpotent orbit is compatible with (\ref{nilc}). The theory flows in the IR to $H_0$. 
\item For $E_7$ MN $D=4$ and the nilpotent orbits compatible with (\ref{nilc}) are the principal and the $E_6$ orbit. The theory flows to $H_0$ and $H_1$ respectively. 
\item For $E_6$ MN $D=3$ and our argument selects the orbits $D_4$, $D_5$ and $E_6$ (i.e. the principal). The theory flows in the IR to $H_2$, $H_1$ and $H_0$ respectively. 
\item For $SU(2)$ SQCD with 4 flavors $D=2$ and our procedure selects the principal nilpotent orbit, the orbits $(5,1^3)$, $(4,4)^I$, $(4,4)^{II}$ and $(3^2,1^2)$. The infrared fixed point is $H_0$ in the first case, $H_2$ in the last and $H_1$ in the other three. 
\item For $H_2$ and $H_1$ all the nilpotent orbits pass our test. The infrared fixed points are $H_1$ or $H_0$. 
\end{itemize}
In all the above cases equation (\ref{an44}) is indeed satisfied. All other rank one theories discussed in \cite{Argyres:2016xmc} do not pass the first step of our criterion and therefore do not exhibit enhancement of supersymmetry in the IR.

\subsection{A test for the existence of a lagrangian UV completion}\label{ciaociao}

We start by recalling that in the infrared the scaling dimension of singlets and CB operators of the UV theory which do not decouple (and therefore become CB operators of the IR SCFT) reads $\frac{3}{2}(1+\epsilon_*)k$. So, if we sum over $k$ for all singlets and all UV CB operators and multiply everything by $\frac{3}{2}(1+\epsilon_*)$, we find the formula  
\be\label{keyformula}\frac{3}{2}(1+\epsilon_*)\left(4a-2c+\frac{r}{2}+\sum_i(s_i+1)\right)=4a'-2c'+\frac{r}{2}+\frac{3}{2}(1+ \epsilon_*)\sum_{j,dec}k_j.\ee
On the l.h.s. we have exploited (\ref{cbformula}) for the UV theory and on the r.h.s. we have rewritten the sum in terms of IR CB operators and decoupled fields, and then used (\ref{cbformula}) again.

Combining (\ref{anmatch1}) and (\ref{keyformula}) we can easily get the relation 
\be\label{key2}\frac{3}{2}(1+\epsilon_*)(40a-44c-r)=40a'-44c'-r,\ee 
and plugging this in (\ref{anmatch}) we find the formula 
\be\label{keyformula2}(6c'-r)(4a-5c)=(6c-r)(4a'-5c').\ee 
The interesting feature of this equation is that it provides a universal relation between the central charges of the UV and IR theories and does not depend on the specific nilpotent orbit considered (apart from the implicit constraint that supersymmetry enhances in the infrared). Even more surprisingly, it does not depend on the global symmetry of the UV theory. 

A nice consistency check of (\ref{keyformula2}) is provided by rank one theories: as we have already said, it was found in \cite{Maruyoshi:2016aim} that all the theories realized by a D3 brane probing a stack of 7-branes ($E_N$ MN theories, $D_4$, $H_2$, $H_1$ and $H_0$) can flow to $H_0$ (under a principal nilpotent vev). By plugging in (\ref{keyformula2}) the central charges for $H_0$ ($a'=43/120$ and $c'=11/30$) we find the relation $12a-9c=1$, which should hold for all theories flowing to $H_0$ and indeed this is true for all the above-mentioned models. As a consistency check we will show in Appendix \ref{checklit} that (\ref{keyformula2}) is satisfied by all known RG flows with supersymmetry enhancement in the IR.  

As the previous discussion shows, equation (\ref{keyformula2}) can be used to provide an a priori constraint on the central charges of a putative UV completion of a given SCFT. In particular, it is rather effective in constraining  lagrangian UV completions of a given SCFT, especially for low-rank theories. In order to illustrate this point it is convenient to use, instead of a and c, the parameters $n_v$ and $n_h$ defined as follows (see \cite{Gaiotto:2009gz}): 
\be\label{nmultiplets}a=\frac{5n_v+n_h}{24};\quad c=\frac{2n_v+n_h}{12}\longrightarrow 8a-4c=n_v;\quad 5c-4a=\frac{n_h}{4}.\ee 
These are the ``effective'' number of vector multiplets and hypermultiplets respectively and coincide with the actual number of multiplets in the case of lagrangian theories.  

The idea is rather simple: because of the rank condition, we know that the lagrangian UV completion should have the same rank as the given SCFT. This tells us the value of the rank of the gauge group, leaving just a finite number of possibilities. Since in the case of lagrangian theories $n_v$ is just the dimension of the gauge group, imposing (\ref{keyformula2}) for all the groups consistent with the rank condition allows us to determine the value of $n_h$ in each case. 

Let's discuss a couple of examples: we first revisit the simple case of Minahan-Nemeschansky theories to see what (\ref{keyformula2}) adds to the argument based on the rank condition we have given previously. The putative UV lagrangian completion should have rank one, meaning the gauge group is $SU(2)$ as we have remarked before. Imposing (\ref{keyformula2}) and setting $n_v=3$ we find $n_h=8$, thus ruling out a priori $SU(2)$ $\CN=4$ SYM as a candidate UV lagrangian completion.  

A more interesting example is the $E_6$ AD theory: the model has rank three so we have eight possibilities for the gauge group of the UV theory. Plugging in (\ref{keyformula2}) the central charges of $E_6$ AD theory ($a'=\frac{75}{56}$ and $c'=\frac{19}{14}$) we find that the UV lagrangian theory has to satisfy the relation $n_h=5n_v/2-15/2$, giving the following list of possibilities: 
\be\label{tabg}\begin{array}{|c|c|c|}
\hline 
\text{Gauge Group} & n_v & n_h \\
\hline 
SU(4) & 15 & 30 \\
\hline 
USp(6) & 21 & 45 \\
\hline 
SO(7) & 21 & 45 \\
\hline 
SU(3)\times SU(2) & 11 & 20 \\
\hline 
USp(4)\times SU(2) & 13 & 25 \\
\hline 
SO(5)\times SU(2) & 13 & 25 \\ 
\hline 
G_2\times SU(2) & 17 & 35 \\
\hline 
SU(2)^3 & 9 & 15 \\ 
\hline 
\end{array}\ee
The only lagrangian SCFT's compatible with these data are the linear quiver 
$$\boxed{1}-SU(2)-SU(3)-\boxed{4}$$ 
and the $SU(4)$ theory with 6 fundamentals and one antisymmetric. Neither of these models flows to $E_6$ AD theory and therefore we conclude that there is no UV lagrangian completion for this theory, at least with an RG flow of this type. 

Actually, we can simplify our test exploiting the remark we have made just after (\ref{1to1}): the number of CB operators of smallest dimension is the same in the IR and in the UV. This observation is useful because in the UV lagrangian theory the CB operators of smallest dimension have dimension 2 and are in one-to-one correspondence with the simple factors in the gauge group. Since the $E_6$ AD theory has  CB operators of dimension $8/7$, $9/7$ and $12/7$, we conclude that the UV lagrangian theory should have just one operator of dimension 2 and this restricts the list of candidate gauge groups to the first three appearing in (\ref{tabg}); in particular this argument rules out automatically the $SU(2)\times SU(3)$ quiver. 

The fact that the $SU(4)$ theory does not flow in the IR to $E_6$ AD theory  can be checked for example using a-maximization, but there is also a faster argument which allows us to narrow down further the list of candidate UV lagrangian completions: as we have already explained, the scaling dimension of CB operators in the IR is $\frac{3}{2}(1+\epsilon_*)k$ (using the parameter $k$ defined in \ref{kpara}) and one can easily check whether a value of $\epsilon_*$ exists such that the candidate UV theory reproduces the spectrum of CB operators of the given SCFT. Let's see how this works in the example at hand: considering e.g. the next-to-maximal $SU(6)$ nilpotent vev for the $SU(4)$ theory, the candidate CB operators in the IR have $k=4,4,5$ and clearly there is no value of $\epsilon_*$ which reproduces the spectrum of $E_6$ AD theory. The principal nilpotent vev is not compatible with the rank condition and therefore is ruled out a priori. 

It is worth pointing out that the number of nilpotent vevs to be checked can be reduced exploiting the fact that supersymmetry enhancement in the infrared can occur only if there is at least one singlet with charge under $R_{\CN=2}-2\rho(\sigma_3)$ larger than all UV CB operators. For example, in the case at hand only the principal and next-to-maximal nilpotent vevs are compatible with this constraint and we have already explained why they do not work.

\section{Concluding remarks}

In this note we have shown that, in order to have supersymmetry enhancement in the infrared, basic properties of $\CN=2$ SCFT's and 't Hooft anomaly matching impose stringent constraints on the RG flow and especially on the UV fixed point. This also allows us to put strong constraints on a lagrangian UV completion of a given $\CN=2$ SCFT. One of the most interesting outcomes of our analysis is equation (\ref{an441}), which characterizes SCFT's which exhibit, under the Maruyoshi Song flow, enhancement of supersymmetry in the infrared. In \cite{Agarwal:2017roi} it was proposed that the criterion for supersymmetry enhancement can be formulated in terms of the corresponding chiral algebra \cite{Beem:2013sza} and it would be intersting to find an interpretation of  (\ref{an441}) along these lines. 

As was mentioned earlier, the analysis of  \cite{Agarwal:2016pjo}-\cite{Benvenuti:2017bpg} strongly suggests the following fact: whenever the theory does not flow to an IR fixed point with extended supersymmetry under a principal nilpotent vev, other choices of nilpotent vev will not work either. This property does not seem to be directly implied by our analysis and it would be interesting to elucidate further this point.   

In principle our approach can be adapted to study supersymmetry enhancement for other types of RG flows as well. The basic requirement is that the Cartan subgroup of the infrared R-symmetry is visible in the UV; this is needed to write down (\ref{rir1}), which constitutes the starting point of our construction. The nontrivial input we need is an a priori guess for the IR CB operators: we should know in advance which chiral operators in the UV theory become CB operators in the infrared. The assumption we described in the introduction, which plays a key role in our analysis, precisely provides this information. We hope this construction will be helpful to identify new examples of RG flows which exhibit supersymmetry enhancement at long distances.

\section*{Acknowledgements} 

I am grateful to Federico Carta and Raffaele Savelli for many interesting discussions and for collaboration on closely related topics. I also acknowledge Sergio Benvenuti and Mario Martone for discussions and comments and especially Philip Argyres and Yuji Tachikawa for carefully reading the manuscript. I would like to thank the Department of Physics of Milano Bicocca and the Galileo Galilei Institute for Theoretical Physics for  hospitality during the completion of this project. This work is partly supported by the INFN Research Project ST$\&$FI.

\appendix 

\section{Comparison with a-maximization}\label{maxcheck}

In this Appendix we are going to prove that the a-maximization analysis is automatically consistent with (\ref{amaxf}) for all theories satisfying our criterion. Let's start by recalling how the a-maximization procedure works (see \cite{Maruyoshi:2016aim, Agarwal:2016pjo}): the trial a central charge is \be\label{amax}a(\epsilon)=\frac{3}{32}(3\Tr R_{\epsilon}^3-R_{\epsilon})=a_{UV\;SCFT}(\epsilon)+a_{\text{singlets}}(\epsilon)-a_{\text{decoupled}}(\epsilon),\ee 
where $R_{\epsilon}$ is the generator (\ref{rtrial}) and $a_{UV\;SCFT}(\epsilon)$, $a_{\text{singlets}}(\epsilon)$, $a_{\text{decoupled}}(\epsilon)$ denote the contributions from the UV SCFT, the singlets and decoupled fields. Explicitly these read 
\be a_{UV\;SCFT}(\epsilon)=\frac{27}{16}(1+\epsilon)^3(a-c)+\frac{27+27\epsilon}{64}[(1-\epsilon)^2(4a-2c) -(1+\epsilon)^2I_{\rho}\beta_G]-\frac{9+9\epsilon}{4}(a-c),\ee 
\be a_{\text{singlets}}(\epsilon)=\frac{3}{32}\sum_i\left[3\left((1+\epsilon)\frac{2s_i+1}{2}+\frac{\epsilon-1}{2}\right)^3-(1+\epsilon)\frac{2s_i+1}{2}-\frac{\epsilon-1}{2}\right],\ee 
\be a_{\text{decoupled}}(\epsilon)=\frac{3}{32}\sum_{j,dec}\left[3\left((1+\epsilon)\frac{2k_j-1}{2}+\frac{\epsilon-1}{2}\right)^3-(1+\epsilon)\frac{2k_j-1}{2}-\frac{\epsilon-1}{2}\right].\ee 
Notice that exploiting the relation 
$$R_{\epsilon_*}=\frac{1}{3}R'_{\CN=2}+\frac{4}{3}I'_3$$ 
and using (\ref{anmatch1}),(\ref{anmatch2}),(\ref{anr2}),(\ref{an3final}) we find as expected 
$$a(\epsilon_*)=a'.$$ 
The statement of a-maximization is that the value of $\epsilon$ corresponding to the actual IR R-symmetry maximizes $a(\epsilon)$, so we have to show that at $\epsilon_*$ the first derivative vanishes and the second derivative is negative. 
 
The fact that the derivative of $a(\epsilon)$ evaluated at $\epsilon_*$ vanishes can be seen as follows: if we take the r.h.s. of (\ref{anr3}) and add six times the l.h.s. of (\ref{anr2}) we get of course a vanishing expression however, it is just a matter of straightforward algebra to check that this is also equal to $32\frac{da}{d\epsilon}(\epsilon_*)$. In matching the two expressions we use the fact that in (\ref{amax}) we have set the number of decoupled operators to be equal to the number of singlets.

In order to prove that the second derivative is negative, it is useful to look at its explicit form\footnote{We are dividing everything by an overall factor of 18.}: 
\be\label{secder}
  4a-5c -\frac{3}{2}\beta I_{\rho}+\sum_is_i(s_i+1)^2 -\sum_{jdec}k_j^2(k_j-1)+ 
   \epsilon\left(12a-9c-\frac{3}{2}\beta I_{\rho}+\sum_i(s_i+1)^3-\sum_{jdec}k_j^3\right)\ee
We can notice that this expression simplifies considerably at $\epsilon=-1$ and reduces to 
$$-(8a-4c)-\sum_i(s_i+1)^2+\sum_{jdec}k_j^2.$$ 
This quantity is manifestly negative: $\sum_i(s_i+1)^2-\sum_{jdec}k_j^2$ is clearly positive (see also \ref{cnuova}) and $8a-4c$ is positive due to the Maldacena-Hofman bound \cite{Hofman:2008ar}. We therefore learn that the second derivative is negative at $\epsilon=-1$. Since (\ref{secder}) is linear in $\epsilon$, in order to conclude the argument it is enough to look at the point $\tilde{\epsilon}$ where the second derivative vanishes and keep in mind the fact that  $\epsilon_*>-1$ because of (\ref{singc}).  
From (\ref{secder}) we have\footnote{If the coefficient of the linear term in (\ref{secder}) vanishes the formula for $\tilde{\epsilon}$ does not hold. However, in this case the second derivative is a constant and since it is negative at $\epsilon=-1$ it will be negative at $\epsilon_*$ as well.} 
\be\frac{3}{2}(1+\tilde{\epsilon})=\frac{3\left(8a-4c+\sum_i(s_i+1)^2-\sum_{jdec}k_j^2\right)}{24a-18c-3\beta I_{\rho}+2\sum_i(s_i+1)^3-2\sum_{jdec}k_j^3},\ee 
and we recognize at the numerator and at the denominator the expressions appearing in (\ref{anr2}) and (\ref{an3final}) respectively. Using these equations we therefore conclude 
\be(1+\tilde{\epsilon})=\frac{3}{2}(1+\epsilon_*)\frac{8a-4c+\sum_i(s_i+1)-\sum_{jdec}k_j}{12a-9c+\sum_i(s_i+1)-\sum_{jdec}k_j}.\ee 
If the denominator in this equation is negative we conclude that $\tilde{\epsilon}<-1$ and since the second derivative is linear in $\epsilon$ and is negative at $\epsilon=-1$, it is necessarily negative at $\epsilon=\epsilon_*$ as well (just because $\epsilon_*>-1$). If instead the denominator is positive, we need to show that 
$$\frac{8a-4c+\sum_i(s_i+1)-\sum_{jdec}k_j}{12a-9c+\sum_i(s_i+1)-\sum_{jdec}k_j}>\frac{2}{3}.$$ 
This inequality is just equivalent to 
$$\sum_i(s_i+1)-\sum_{jdec}k_j+6c>0,$$ which is manifestly true.

\section{Supersymmetry enhancement and Type IIB}\label{checklit}

In this Appendix we are going to compare our findings with results available in the literature.
In \cite{Giacomelli:2017ckh} it was noticed that all examples of infrared supersymmetry enhancement found so far can be embedded in Type IIB string theory, so it is enough to check (\ref{keyformula2}) in that context. The class of models we are interested in is engineered by compactifying Type IIB on local three-folds of the following form (see \cite{Wang:2015mra}):
\be\label{singol}\begin{array}{|c|c|c|}
\hline
J & \text{Singularity} & b \\
\hline 
A_{N-1} & x_1^2+x_2^2+x_3^N+z^k=0& N \\
\hline
& x_1^2+x_2^2+x_3^N+x_3z^k=0& N-1 \\
\hline 
D_N & x_1^2+x_2^{N-1}+x_2x_3^2+z^k=0& 2N-2 \\
\hline 
 & x_1^2+x_2^{N-1}+x_2x_3^2+x_3z^k=0& N \\
\hline 
E_6 & x_1^2+x_2^3+x_3^4+z^k=0& 12 \\
\hline
 & x_1^2+x_2^3+x_3^4+x_3z^k=0& 9 \\
\hline
 & x_1^2+x_2^3+x_3^4+x_2z^k=0& 8 \\ 
\hline 
E_7 & x_1^2+x_2^3+x_2x_3^3+z^k=0& 18 \\
\hline
 & x_1^2+x_2^3+x_2x_3^3+x_3z^k=0& 14 \\ 
\hline 
E_8 & x_1^2+x_2^3+x_3^5+z^k=0& 30 \\
\hline 
 & x_1^2+x_2^3+x_3^5+x_3z^k=0& 24 \\
\hline 
 & x_1^2+x_2^3+x_3^5+x_2z^k=0& 20 \\
\hline
\end{array}\ee
These are all hypersurfaces of the form $\mathcal{W}(x_1,x_2,x_3,z)=0$ obtained by fibering an ADE singularity (parametrized by $x_{1,2,3}$) on the $z$ plane. The data of the fibration is encoded in the parameter $k$ (which is an arbitrary positive integer) and the integer $b$ as in (\ref{singol})\footnote{When $b$ is equal to the Coxeter number of $J$, we recover the models studied in \cite{Cecotti:2010fi, Cecotti:2013lda}}. If $z$ is a coordinate in $\mathbb{C}^*$ the resulting models were called $D_k^b(J)$ in \cite{Giacomelli:2017ckh} and have global symmetry (at least) $J$. If instead $z$ is $\mathbb{C}$-valued we find theories called $J^b(k)$ which in general do not have any global symmetry\footnote{The main difference between these two classes is the normalization of the holomorphic three-form, which fixes the scaling dimension of BPS operators in the theory.}. 
The observation of \cite{Giacomelli:2017ckh} is that the $D_k^b(J)$ theory flows in the IR to $J^b(k)$ under a principal nilpotent vev for the $J$ adjoint chiral. 

Let's start by checking (\ref{amaxf}): in the case of $D_k^b(J)$ theories $D_{max}=h-b/k$ and $s_{max}=h-1$, therefore we have 
\be\label{cccc} \frac{3}{2}(1+\epsilon_*)=\frac{1}{2+s_{max}-D_{max}}=\frac{k}{k+b},\ee  
which agrees precisely with the value of $\epsilon_*$ guessed in \cite{Giacomelli:2017ckh}. Our construction can be seen as a derivation of that result.

We now proceed by noticing that, if we combine (\ref{anmatch}) with (\ref{keyformula2}), we find the relation 
\be\label{epsilon}\frac{3}{2}(1+\epsilon_*)=\frac{6c'-r}{6c-r}.\ee
We can now exploit the Shapere-Tachikawa formula \cite{Shapere:2008zf}
\be\label{cb1} 6c-r=2R(B),\ee 
where $R(B)$ is the R-charge of the discriminant, and rewrite (\ref{epsilon}) in the form 
\be\label{cb2} \frac{3}{2}(1+\epsilon_*)=\frac{R(B)^{IR}}{R(B)^{UV}}.\ee 
Using the known formula expressing $R(B)$ for $D_k^b(J)$ and $J^b(k)$ theories (see \cite{Giacomelli:2017ckh, Xie:2015rpa}) we indeed recover (\ref{cccc}). In order to conclude the argument, we now define as in \cite{Shapere:2008zf} 
$$R(A)\equiv\sum_i(D_i-1)$$ 
and rewrite (\ref{keyformula}) as follows: 
\be \frac{3}{2}(1+\epsilon_*)(R(A)^{UV}+r+\sum_i(s_i+1))=R(A)^{IR}+r+\frac{3}{2}(1+\epsilon_*)\sum_{jdec}k_j.\ee 
Using now (\ref{cb2}), this can equivalently be written as 
\be\label{cb3} R(A)^{IR}-R(B)^{IR}=\frac{3}{2}(1+\epsilon_*)(R(A)^{UV}-R(B)^{UV})-r+\frac{3}{2}(1+\epsilon_*)(r+\sum_i(s_i+1)-\sum_{jdec}k_j).\ee 
If we now notice that $$4a-5c=R(A)^{UV}-R(B)^{UV};\quad 4a'-5c'=R(A)^{IR}-R(B)^{IR},$$ 
we can easily see that (\ref{keyformula2}) holds if and only if the sum of the last two terms on the r.h.s. of (\ref{cb3}) vanishes. This is equivalent to the equation 
\be \frac{3}{2}(1+\epsilon_*)=\frac{r}{r+\sum_i(s_i+1)-\sum_{jdec}k_j},\ee 
which is nothing but (\ref{amaxf}) if we plug (\ref{idesimp}) in the above formula. This proves that (\ref{keyformula2}) holds for all the RG flows in this class. 

We can actually explicitly check the validity of (\ref{idesimp}) and more in general (\ref{1to1}) in the case of $D_k^b(J)$ theories: the idea is that the one-to-one correspondence between UV and IR CB operators defined in Section \ref{sec4} has a simple geometric realization in this class of theories. 

In order to explain how this works, we recall that for $D_k^b(J)$ theories the versal deformations of the ADE singularity are the mass Casimirs of the $J$ global symmetry. The vev of UV CB operators is instead described by the $z$-dependent deformation terms. The $J^b(k)$ theory instead is described by the same equation $\mathcal{W}(x_1,x_2,x_3,z)=0$ in $\mathbb{C}^4$ rather than $\mathbb{C}^3\times\mathbb{C}^*$ and the CB operators correspond to all deformation terms with coefficient of dimension larger than one. We can now easily describe the one-to-one correspondence between UV and IR CB operators: given any UV CB operator $u$, divide the corresponding deformation term by $z$. This operation maps the original term to another deformation and the scaling dimension of the corresponding parameter $u'$ is that of $u$ plus the dimension of $z$, which in the $D_k^b(J)$ theory is equal to $\frac{b}{k}$. Since by assumption $D(u)>1$, we conclude that $$D(u')>1+\frac{b}{k}=\frac{k+b}{k}.$$ 
Now we exploit (\ref{cccc}), which tells us that the scaling dimension in the IR of $u'$ (provided it does not decouple) is $D(u')$ times $\frac{k}{k+b}$, and due to the above inequality, we clearly see that this quantity is larger than one. We then conclude that the term $u'$ always corresponds to a CB operator of the IR theory $J^b(k)$. Notice that from (\ref{cccc})
$$-\frac{1+3\epsilon_*}{3+3\epsilon_*}=\frac{b}{k},$$ 
therefore we precisely recover the one-to-one correspondence (\ref{1to1}).

\bibliographystyle{ytphys}

\end{document}